# Efficient Access of Mobile Flows to Heterogeneous Networks under Flash Crowds

Jose Moura
Instituto Universitário de Lisboa (ISCTE-IUL)
Instituto de Telecomunicações
Portugal
jose.moura@iscte.pt

Christopher Edwards
School of Computing and Communications
Lancaster University
United Kingdom
ce@comp.lancs.ac.uk

*Abstract*— **Future wireless networks need to offer orders of magnitude more capacity to address the predicted growth in mobile traffic demand. Operators to enhance the capacity of cellular networks are increasingly using WiFi to offload traffic from their core networks. This paper deals with the efficient and flexible management of a heterogeneous networking environment offering wireless access to multimode terminals. This wireless access is evaluated under disruptive usage scenarios, such as flash crowds, which can mean unwanted severe congestion on a specific operator network whilst the remaining available capacity from other access technologies is not being used. To address these issues, we propose a scalable network assisted distributed solution that is administered by centralized policies, and an embedded reputation system, by which initially selfish operators are encouraged to cooperate under the threat of churn. Our solution after detecting a congested technology, including within its wired backhaul, automatically offloads and balances the flows amongst the access resources from all the existing technologies, following some quality metrics. Our results show that the smart integration of access networks can yield an additional wireless quality for mobile flows up to thirty eight percent beyond that feasible from the best effort standalone operation of each wireless access technology. It is also evidenced that backhaul constraints are conveniently reflected on the way the flow access to wireless media is granted. Finally, we have analyzed the sensitivity of the handover decision algorithm running in each terminal agent to consecutive flash crowds, as well as its centralized feature that controls the connection quality offered by a heterogeneous access infrastructure owned by distinct operators.**
   *Heterogeneous Access, WiNeMOs, Congestion, Distributed Management Algorithm, Embedded Reputation, Mobile Flows*

1. INTRODUCTION

According a recent traffic forecast, IP will increase at a Compound Annual Growth Rate (CAGR) of 23 percent from 2014 to 2019; from this growth the mobile data has a very significant share. Additionally, busy-hour (i.e. the busiest 60 minute period in a day) Internet traffic is growing more rapidly than the average. In fact, busy-hour Internet traffic increased 34% in 2014 and, during the same year, the average traffic increased only 26%. It is expected that the busy-hour Internet traffic will increase by a factor of 3.4 between 2014 and 2019, while the average will only increase 2.8-fold [1]. It is also estimated that the terminal devices will change very quickly. In fact, during the period (2014, 2019), PC-originated traffic will grow at a CAGR of 9%, while TVs, tablets, smartphones, and Machine-to-Machine (M2M) modules will have traffic growth rates of 17%, 65%, 62%, and 71%, respectively.

Following previous predictions, we argue that the available access capabilities of any heterogeneous network should be more efficiently used to attenuate the congestion problem due to flash crowds. The current contribution follows on previous work about a novel functionality to manage the rate of some selected flows, depending on the available network resources and load characteristics [2]. Despite the huge progress in terms of interworking research [3], due to the recent evolution on the network requisites and its heterogeneous accessibility, further work is needed to control mobility, flow admission and network selection. These aspects are completely revisited by the current contribution, which proposes a complete solution amongst diverse access technologies based on a reputation based-system that guarantees a specific Quality of Service (QoS) level per flow. The flow QoS is administered by our proposal that uses an innovative and dynamic status aggregation of both wireless media and backhaul access. In addition, the QoS flow is entirely managed by networking policies and a set of convenient quality metrics based on the past and current performance of access technologies.

The major contributions of the current paper are six-fold as follows. First, a planning tool was used to analyze the long-term impact of deployment strategies in a heterogeneous network infrastructure on both financial and performance aspects, using a brokerage system. Analyzing the results from the last tool, one can conclude that a broker-based solution enhances the wireless quality offered by the access technologies in the presence of flash crowd situations. It also creates more fairness in the market, rewarding in financial terms the provider who invested more money on upgrading its network. Second, we show that using a hybrid system with the right policies applied in a centralized way over distinct wireless access technologies, it is possible to keep all the accepted flows with a maximum quality value in spite of very high loads, using also a completely distributed management algorithm. Third, potential backhaul bottlenecks are also efficiently detected and solved by intelligently controlling new flow admission and/or offloading traffic amongst access technologies. Fourth, the low-complexity distributed solution supervises mobile terminals correctly, as a correct set of policies is also deployed. Fifth, the solution is load-aware in that it balances in a stable way the traffic load on the diverse wireless and wired links, increasing the network global throughput by routing flows through less congested paths over heterogeneous resources, enhancing flow quality. Finally, the current proposal operates in a very similar way to the emerging Software Defined Networking (SDN) paradigm.

This paper has a structure as follows. Section 2 contextualizes the current contribution in the literature. Section 3 describes major functional requirements of our proposal. Section 4 discusses comprehensively its design. Section 5 analyzes the long-term advantages obtained by operators when our distributed proposal manages a heterogeneous infrastructure using a real network loading. Section 6 discusses how the solution was implemented, and Section 7 evaluates it as proof of concept. Finally, Section 8 concludes our work.

2. RELATED WORK

The literature was reviewed, highlighting the novel aspects of our work, in the following aspects: i) finding the most effective supervision mechanisms to offer a good connectivity service through a heterogeneous access network with high traffic load; ii) identifying the more suitable ways to deploy the previous mechanisms, such as brokers or SDN; iii) how to implement management policies, some of these based on the reputation metric, to control the heterogeneous networking capabilities as efficiently as possible.

*2.1. Media Independent Access*

There are some open problems in the successful integration of different wireless access technologies [4] to supply a media independent access to mobile devices [5]. We aim to address some of these problems with our current cross-technology proposal (Table 1): flow admission, network selection, load balancing, deployment of a low-complexity, flexible, and sharing-focused solution, sharing resources among networks owned by distinct operators. To tackle these issues, our novel supervision proposal, for each technology, measures the status of both wireless and backhaul access links and evaluates several QoS metrics, including the reputation, depending on specified weights. Then, the proposal disseminates periodically all these metrics amongst all the existing terminals. In addition, each terminal supports a novel and dynamic cost function to rank target Network Attachment Points (NAPs) based on a variety of metrics, such as NAP quality, terminal signal power received and technology (i.e. operator) reputation. Other metrics could be easily added like the connection cost, security, error rate, or latency. Our quality metrics are load-aware, very similar to the Load Aware Routing Metric [6] applied to multi-hop networks but for a distinct scenario. In fact, the former was for a single technology and ours is for a heterogeneous system.

The terminals ranked list of NAPs, eventually from different access technologies, in combination with convenient policies downloaded from the brokerage service deployed at the access network, enhance at the terminals, very important control features like flow admission and network selection, which are, in this way, executed at terminals but assisted by the network (Table 1). Our proposal is related with available contributions in [7, 8]. Nevertheless, our proposal has a broader utilization than [7] because the latter is specialized to offload voice traffic. In opposition to [8], where a centralized solution is proposed to optimize the network access in scenarios with no mobility support, the current work offers a distributed flow admission mechanism to support a specific QoS flow level in each multimode mobile terminal. Our solution also supports load balancing among wireless access technologies [9] and traffic offloading [7]. In this way, the flow handovers are not only due to terminal mobility but also to load variations at NAPs and/or backhaul, which are logically controlled by policies in a centralized way, similarly to the SDN paradigm (Table 1).

We have deployed the distributed control algorithm to supervise the handovers of mobile multimode terminals within the terminals themselves because in this way we reduce the complexity, signaling overhead and handover latency [10-12]. The mobility management is comprehensively surveyed in [13] and, other work [14] evaluates analytically the handover delay when common mobility protocols are deployed in heterogeneous networks using IEEE 802.21 [15]. The handover delay is not relevant in our work because we assume make-before-break handovers using the macro-mobility protocol MIPv6. In addition, our mobility proposal is a Network-Assisted Mobile terminal Initiated HandOver (NAMIHO) [16].

Table 1 – Summary of new functionalities supported by our current proposal

| Addressed Problem | New Functionality | Main Goal |
|---|---|---|
| Flow admission among heterogeneous access technologies | Policies made available to terminals from the brokerage service in execution at the network layer in different network nodes | To guarantee a specific QoS level (compatible with existing quality standards) to an individual flow through all the access technologies. Otherwise the flow is blocked. |
| Network selection among technologies | Metrics received from the network together with policies allow terminals to create dynamically a ranked list of available NAPs | Terminal assisted by network uses a novel cost function to select dynamically the most suitable NAP, always reflecting the status of wireless and backhaul parts; it reduces latency and network overhead |
| Load balancing and traffic offloading among technologies | Flow handover is triggered at the terminal side by high traffic loads at wireless media and/or backhaul | In spite of severe loads, our proposal efficiently uses the available network capabilities (backhaul and wireless) from all the technologies, fulfilling flows quality |
| Deploy a low-complexity and flexible management to support any heterogeneous access architecture | Hybrid bandwidth brokerage service implemented by software routines in strategic network nodes and terminal agents, all running at the network layer | The proposed broker system applies policies in a centralized way and enables technologies to be controlled in a distributed way; it operates similarly to SDN |
| Sharing of network capacity among technologies | Reputation is used in the user admission and network selection features | Operators are forced to cooperate, sharing available resources, keeping their reputation and profit as high as possible |

## 2.2. Brokers

Using brokers to control networks has been extensively investigated [2, 17-19]. Other more recent paradigms such as SDN and Network Functions Virtualization (NFV) [20-22] are also related to our current work. The broker nodes of our solution can be seen as SDN controllers. In addition, the abstracted vision of the physical network topology ensured in our case by 802.21 has a very strong similarity with the NFV abstraction controlled by high-level management policies. Therefore, our proposal with its main innovations can be used into a SDN/NFV solution to control in an efficient way networking access infrastructures with higher levels of complexity, diversity, dynamics, scale and virtualization. The joined operation of SDN and 802.21 has been evaluated in a real testbed to optimize handovers within a wireless environment [23]. NFV management and orchestration challenges are discussed in [22].

## 2.3. Reputation as a cooperation mechanism

Previous work applied several mechanisms to incentivize the cooperation within networks [24-29]. These mechanisms can be classified in two major approaches, credit-based [24] and reputation-based [25-28]. The authors of [24] propose a credit-based system for stimulating cooperation among selfish nodes in mobile Ad-Hoc networks. Alternatively, reputation was used in Ad-Hoc [25, 27] and wireless sensor [26] networks to encourage a cooperative behavior. Game theory has been also recently surveyed [28] to understand how reputation can enhance the collaboration among players to use the available resources of Future Networks in a more efficient way. Nevertheless, reputation systems may have vulnerabilities in which malicious attackers can exploit (collude) to perform sequences of misleading behavior with the aim of gaining unfair trust/reputation scores. More on this is available in [29].

Our work is a reputation-based system that aggregates the available network resources of both backhaul and wireless access links in a single access infrastructure, making it available to multimode mobile terminals. The heterogeneous network (scarce) capabilities are effectively controlled as they belong to the same technology. We explain now briefly how this is made. It is used a reputation system that grants a wireless network access to multimode terminals in a distributed way based on the past behavior of each available access technology [30]. A good behavior results in escalation (incentive) of the number of terminals connected to the technology, whereas a bad behavior implies less (punishment) terminals be attached to that technology. In this way, past behavior has a direct impact on the operator profit, naturally forcing operators to keep their network reputation as high as possible. Consequently, operators should cooperate [28, 31], enhancing the utilization of the available network resources and increasing their long-term profits. So, network resources are more efficiently managed by policies (i.e. threshold decision metrics including the one related to reputation) that enable vertical handovers of flows or the flow multipath routing (i.e. sub-flows) among the technologies, always transparently to users.

### 3. SCENARIO AND REQUIREMENTS

The following text discusses the usage scenario, the tackled problems and the major requisites for a complete proposal to cope with node mobility and congestion. The typical scenario of our work is visualized in Figure 1. It is a public location covered by different wireless technologies. Each access technology is administered by its own operator. The number of mobile terminals (or nodes) in the coverage zone, demanding wireless communications, changes through the day. In this case, a single access technology could be insufficient to guarantee permanently a connection quality with enough QoS / Quality of Experience (QoE) to all attached terminal flows, due to constraints imposed by the physical wireless channel. These channel impairments could be originated by a few reasons such as terminal mobility, distortion, interference, and fading. Consequently, the current brokerage solution, ruled by policies, attempts to offload, balance, or block the traffic load across the available resources of the entire heterogeneous access network, offering a connection with the expected quality level to the end-user mobile terminals and mitigating congestion.

Figure 1 clarifies that the brokerage solution uses a loosely coupled interworking design. This brokerage operation is totally distributed between access technologies and Points of Presence (PoPs). A PoP is a physical location at the edge of the network where the traffic originated from distinct access technologies is aggregated and routed to the Internet via wired backhaul links. The broker controls the network edge on behalf of operators. A peer-to-peer (P2P) coordination could also exist amongst neighboring brokerage services (Figure 1 – link #6). This coordination is very helpful to predict the occurrence of congestion in a certain area using events propagated by other networking domains, reporting for example that a large amount of data demand could be expected very shortly due to users just arriving by train to a specific domain.

Without loss of originality, we have made a few assumptions. First, each customer terminal can become connected to the backbone network through any available access technology. The unique condition to be fulfilled is that each customer should hold a contract with a single operator. We assume there is no additional cost for roaming users because there is a business model among operators to enhance the cooperation among them. In fact, a specific operator can charge other operators, without enough network capacity for a very high load, to compensate the former by the fact that some network resources have been used by customers from others. At the end, if all the cooperative operators shared a similar amount of network resources among them, reaching a financial equilibrium, there is no payment among them. Second, the brokerage service assesses the quality of each communications path formed by a NAP wireless link and the associated wired backhaul, the quality of each access technology and the operator's reputation. Third, the brokerage service proactively informs each NAP with updated metrics referring to the status of the associated local networking infrastructure. Fourth, the customer's terminal is frequently notified with quality metrics from all existing NAPs, before it chooses one. Fifth, each MN can associate quality metrics with the received power signal from each NAP to rank that NAP. After this, the MN can finally select and initiate a connection request to the top-of-rank NAP. The distributed management proposal just discussed offloads, balances, or controls the network access of high amounts of load, among the available connection capacity offered by all the technologies, in response to flash crowds and, using general or technology-specific policies.

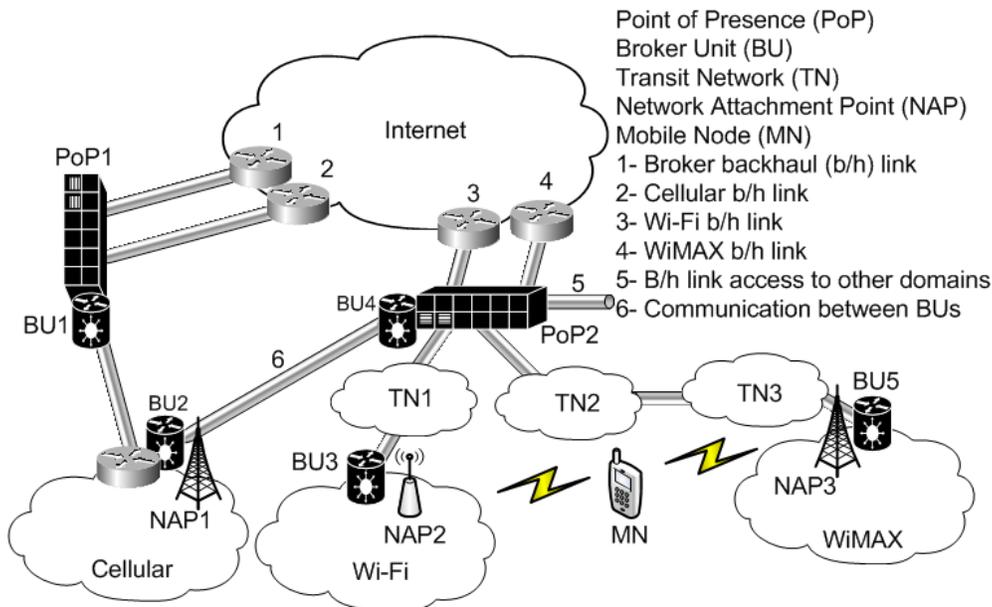

Figure 1 - Scenario of a heterogeneous network infrastructure owned by distinct operators

The brokerage service can be also applied to other network topologies and business models. In the case of a wide area scenario with a very large number of NAPs of the same technology, our solution can easily scale to a hierarchical architecture of broker units. At the bottom level of this architecture, the large number of NAPs can be divided in diverse small-size clusters. Inside each cluster the NAPs geographically close can be grouped and directly controlled by a single broker unit. Then, all these (lowest level) units communicate with other brokers located at the level immediately above. In this way, the status of the technology can be gradually aggregated in a level-by-level way up to reach the top-level aggregating broker unit. Our proposed service can also administrate a cellular topology formed by femtocells. Typically, each femtocell is connected to the operator's core network via a tunnel over a customer´s broadband connection. In this scenario, assuming the existence of a broker unit at each tunnel side, customer flows can be smartly routed and balanced between macrocell and a picocell; it can also mitigate the interference across dense and un-coordinated deployments of femtocells to optimize the downlink performance [32]. Another interesting scenario is the deployment of a Mobile Virtual Mobile Operator without any infrastructure. Finally, our service could support new business models such as Consumer-centric Business Model [33].

4. SOLUTION DESIGN

We discuss the design of our proposal, as follows: Section 4.1 presents our modular, flexible, and scalable solution; then, we go through Sections 4.2-4.4 deeper inside each node's architecture; Section 4.5 describes our vision to deliver end-to-end quality to flows; and Section 4.6 discusses quality metrics and policies.

*4.1. Global Vision of the Brokerage Service*

We explain now the functionality of our solution, using a scenario formed by 2 broker slave units (for the sake of simplicity, the broker master unit was omitted), 2 NAPs (one per access technology) and 1 client's agent operating in a multimode mobile terminal. Nevertheless, one should note that our solution can easily scale in distinct ways: number of providers, technologies owned by each provider, NAPs of each technology, and terminals. Our solution is also very flexible because it can control either small or wide network domains, and it is completely independent of the access technologies being used.

In our scenario, a reason for load increasing is due to the arrival of new customers to a public area covered by both wireless access technologies. The customers' multimode terminal receive simultaneously broadcast link layer messages via distinct technologies and after they select the best technology. Each broadcast message transports in piggyback some quality metrics of the associated technology. Alternatively, these broadcasts can be sent via a Cognitive Pilot Channel [34].

Then, the client's agent processes the received information to choose the available technology with the best quality. One should note that for more complex scenarios with several NAPs per technology and high loads, the terminal's agent might have more options (than accept/block) to control the new flow, such as load balancing in either a horizontal (within the same technology) or vertical (between distinct technologies) way.

The final goal of this distributed management proposal is to keep the quality of each connection as high as possible to avoid churn of customers. Our design will be discussed as follows: broker nodes (4.2), NAP (4.3), terminal's agent (4.4), end-to-end quality of service (4.5), and how quality metrics are processed (4.6).

*4.2. Broker Nodes*

Returning again to Figure 1, there are two broker types: master and slave. In this way, $BU_1$ and $BU_4$ are master units, one for each PoP, and the remaining units are slaves, one for each access network (or technology). The master unit is responsible for the following functional aspects: receives the status of distinct technologies from the associated slaves; implements a centralized policy that hierarchizes the technologies consigning a distinct priority to each one; and updates slaves with the new priority values.

Each broker slave unit is associated with a single access technology (and a unique service provider with an associated reputation metric), evaluating some metrics: backhaul status, each NAP quality, and the technology quality. A detailed explanation how these metrics are calculated is as follows. Firstly, the slave unit assesses the backhaul status of that technology. It actively measures the Round Trip Time (RTT) to evaluate the backhaul load, exchanging periodic ICMP messages with a designated Internet server. This measurement method was used because RTT can be an excellent summary indicator of the performance experienced by diverse flows sharing the same backhaul link [35]. When the measured RTT is below the congestion threshold, the backhaul quality is always one. Alternatively, the slave unit updates the backhaul quality, as in (1), where *RTT* is the measured value. All the parameters/metrics used by our proposal always have a value within the range [0, 1] unless otherwise specified. To achieve this, diverse fitness parameters (i.e. policies) are used. Please refer to Section 4.6 for further details.

$$Q_{back} = \frac{RTT_{max} - RTT}{K} \tag{1}$$

Secondly, the slave unit then combines the backhaul quality (1) with the wireless link quality (see (4) below) received from each NAP, evaluating the quality associated to that NAP, as shown in (2).

$$Q_{NAP} = w_1 * WQ_{NAP} + w_2 * Q_{back} \tag{2}$$

Thirdly, the slave unit calculates the quality of a technology (i.e. Reputation of the provider owning that technology) as the average of all the *m* NAPs using that wireless access technology (3). Then, a specific slave unit distributes the quality metrics among associated NAPs and the broker master unit in one of two ways: event-triggered or periodic reports. A slave unit also gets from each associated NAP a periodic or an event-triggered report with its wireless status. The dissemination of relevant metrics amongst broker units, NAPs, and terminals is done via an augmented version of Media Independent Handover (MIH) [15].

$$Reputation_{prov} = Q_{techno} = \frac{1}{m} * \sum_{j=1}^{m} Q_{NAP_j} \qquad (3)$$

*4.3. Network Attachment Point*

The NAP is our generic designation of the networking node that has two interfaces: one is wireless and the other is wired. The NAP participation in the brokerage service is now discussed. Firstly, a NAP evaluates the quality of its associated wireless link measuring the load of that link, i.e. counting the number of attached flows. This implies the wireless NAP quality is evaluated as shown in (4). In this way, as the number of flows (*n_flow*) attached to a specific NAP increases then the wireless quality of that NAP inversely decreases. Secondly, each NAP updates via the wired media the associated slave broker node with the average wireless quality of the NAP's channel. Thirdly, the broker node combines the received average wireless metric with the broker's wired backhaul quality (see (2)), evaluating other metrics (see (3)). Finally, each NAP, after receiving from the associated slave unit the enriched quality metrics, disseminates these via its wireless channel to the terminals' agents (see 4.4). These quality metrics are piggybacked in link-layer messages periodically broadcasted on-the-air by each NAP.

$$WQ_{NAP} = 1 - n\_flow * K_1 \qquad (4)$$

*4.4. Terminal's Agent*

After the link update from a specific interface technology, the terminal's agent at the network layer ranks the quality of the NAP being processed, combining physical RSS (Received Signal Strength), related with the distance between the terminal and that NAP) and NAP quality with distinct weights, as visualized in expression (5). In this multicriteria expression, a reputation value around one means that in the past the technology of the NAP has performed well and hopefully it will continue to perform well, or at least better than other technologies that in the past had lower reputation values. Consequently, between two NAPs of different technologies but with identical values of received power and quality, expression (5) assigns a higher priority position in the ranking list to the one associated to the technology with the highest reputation. This functionality is very important because it supports a distributed reputation-based system, giving incentives to providers to keep their reputations as high as possible in case of congestion, enhancing cooperation among providers and controlling in a very effective way the usage of all the available heterogeneous resources.

$$P_{NAP} = Reputation_{prov} * (\propto * \frac{power - Pow_{Thr}}{Pow_{Thr}} + (1-\propto) * \frac{Q_{NAP} - Qual_{Thr}}{Qual_{Thr}}) \qquad (5)$$

Using the ranked list of available NAPs (eventually from distinct access technologies), each agent running in the local terminals, after discovering a new NAP or an update on any quality metric of an already known NAP, should validate if the serving NAP is still the best one and if its quality is over the Quality Threshold (QT). Otherwise, the agent performs a flow handover to the new best NAP (only if this new best NAP has a quality higher than QT; otherwise the flow is blocked).

We assume no node in our system shows a misleading behavior. Otherwise, this problem that can adversely affect the reputation trustworthiness as discussed in [29]. In the following Section, we discuss the end-to-end quality of service ensured by our framework.

*4.5. End-to-End Quality of Service*

A novelty of our proposal is that the same set of Classes of Service (CSs) can be shared among heterogeneous access technologies, eventually enhancing technologies that by default support a best-effort connection. There are three CSs related to how each one rules the flows data rate. First, assuming the broker has a management policy that only enables CS 0, the flow rate between the end-user device and a NAP can undertake any value up to the maximum capacity of the wireless channel. In the most extreme case, a single terminal can eventually use the entire channel capacity, which is very unfair because others are unlikely to obtain a connection. Second, if the broker is using policy CS 1, the flow rate is restricted by the condition that the accumulated rate of all the flows connected to a NAP cannot decrease the NAP quality below the Quality Threshold (QT). Otherwise, high flow rates with low priority are throttled (e.g. via a token-bucket solution) in such a way the NAP quality increases above QT. Finally, if CS 2 is enabled each flow rate cannot have a value higher than its contractual rate. In the case of a congested technology (wireless or backhaul access), the set of terminal agents that initially detect congestion can control the associated flows to increase the quality of that technology over QT. One should note that if CS 1 is used then it could imply some changes in how some quality metrics are evaluated (e.g. Equation (4) should considerer *queue size* instead *number of flows*). CS 2 is the default class in our work; so the *number of flows* measures the load.

We argue that our brokerage service may also simplify the end-to-end QoS interworking between wireless access technologies owned by distinct providers. In this context, the DiffServ interconnection with our service is now discussed as follows. DiffServ is a scalable QoS framework for providing differentiated services to flow aggregates in a hop-by-hop way. In addition, our supported Per-Hop Behavior (PHB) traffic aggregates are: Expedited Forwarding (EF), Assured Forwarding (AF) and Best Effort (BE). Aligned with this, [2] proposed a hybrid broker to optimize the sharing of 4G resources among a high number of flows belonging to diverse PHB classes. Further information about DiffServ is available in [36].

Figure 2 shows an end-to-end interworking QoS scenario between a MN and a remote server. It covers two distinct physical media: radio and wired. Two vertical QoS mappings are also used. Firstly, mapping 1 in Figure 2 vertically interconnects the brokerage service with DiffServ. On the other hand, for each wireless access technology, the mapping 2 in Figure 2 interrelates that technology with the brokerage system.

IEEE 802.11e [37] is now assumed, but the next discussion is valid for any access technology supporting QoS. Table 2 lists three types of traffic: voice, video and background (e.g. email or web). For each traffic type, there is first a vertical mapping between the 802.11e Access Category (AC) associated to that traffic type and an equivalent broker CS; then, there is a second mapping between that CS and a compatible DiffServ PHB. As an example, in 802.11e, consider voice as the traffic with the highest priority (i.e. AC AC_VO). Consequently, AC_VO is mapped to the broker class with the highest priority (i.e. CS2). Finally, CS2 is mapped to the DiffServ PHB aggregate with the highest priority (i.e. EF). The vertical mappings of other traffic types are similarly obtained.

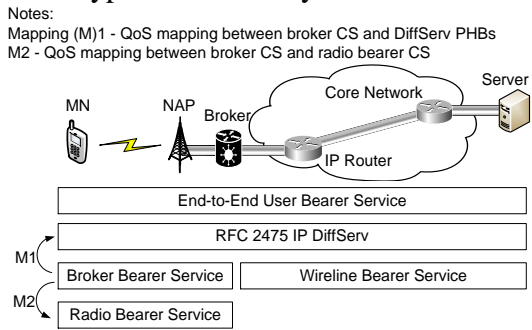

Figure 2 – End-to-End QoS Interworking

Table 2 – Differentiated Classes of Service

| Traffic Type | 802.11e - AC | Broker CS | DiffServ PHB |
|---|---|---|---|
| Voice | AC_VO | CS 2 | EF |
| Video | AC_VI | CS 1 | AF |
| Background | AC_BK | CS 0 | BE |

### 4.6. Quality Metrics and Policies

Figure 3 summarizes the previous discussion about our solution design and visualizes in a holistic way its global operation within a network domain. Figure 3.a) illustrates how the technology status is reported to the brokerage service, and then processed. Figure 3.b) visualizes how the brokerage service after processing the technology status sends a new set of quality metrics to NAPs; how each NAP sends these metrics to terminals; and how terminals agents accordingly refresh their NAP ranking list.

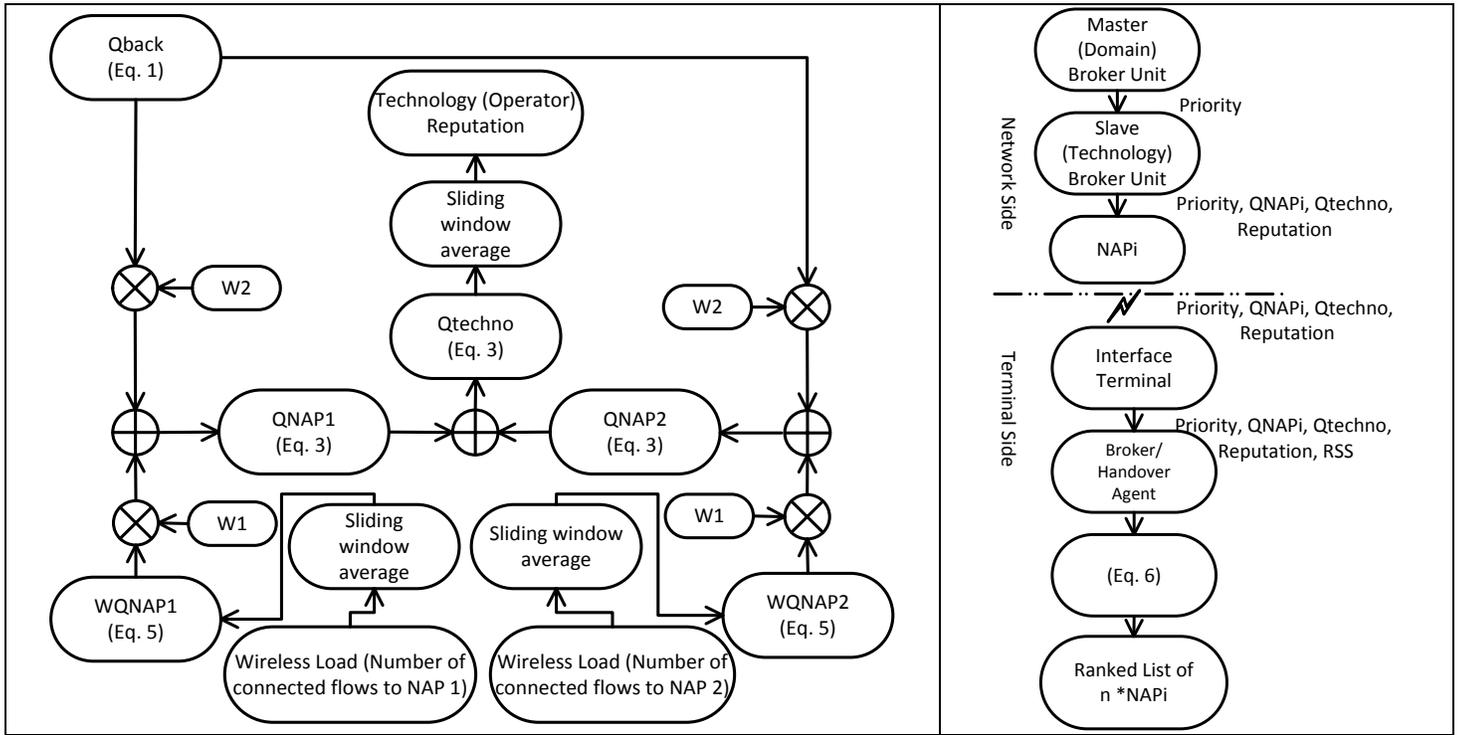

Figure 3.(a)-Quality Metrics for a technology with two NAPs and one B/H link (only Network Side) controlled by a brokerage service

Figure 3.(b)- Brokerage service updates the terminals with Quality Metrics and agents refresh their NAPs list

We now revisit the metrics expressions used in our brokerage system that were already presented in Sections 4.2-4.4. Our intention is in a centralized way to justify the pertinence and give further details about some parameters used in the evaluation of these metrics. To guide this discussion, we introduce the concept of a policy. A policy is a set (tuple) of a variable number of parameters that enables the same metric to be used with distinct access technologies. In this way, the diverse elements of a specific policy behave like fitness parameters to guarantee the metric associated to that policy always returns a value within the range [0, 1], independently of the access technology being monitored and supervised. Table 3 summarizes the most relevant status information that is acquired by the brokerage system, the input variables representing that status, the associated (fitness) policy, and the main functional goal to achieve by processing that status.

Table 3 – Brokerage Service Management Policies

| Status of | Input (Network Status) | Policy = $(p_1, p_2, …, p_n)$ | Goal |
|---|---|---|---|
| Backhaul (B/H) | RTT | $(RTT_{MAX}, K) = (300, 9600)$ | Detects congestion @ B/H |
| NAP (only Wireless) | n_flow | $K_1$=0.0183 (WiMAX); =0.0524 (Wi-Fi) | Detects congestion @ wireless link |
| NAP (B/H+Wireless) | $WQ_{NAP}$, $Q_{back}$ | $(W_1, W_2) = (0.8, 0.2)$ | Aggregates B/H and wireless loads |
| Terminal Agent | Power, QNAP, Reputation, Priority | (Alpha, $POW_{THR}$, $QUAL_{THR}$, delta, QT) = (0.2, 7*10^-9, 0.525, 0.33\|\|0.6, 0.525) | Selects and issues the flow handover to the top viable NAP from a ranked list of options |

The $RTT_{max}$ (i.e. maximum expected value for RTT) and K are scaling parameters to ensure that the backhaul quality is always within the range [0, 1] in case of a congested backhaul link. These parameters should be adjusted to the characteristics of the backhaul access. In our work, we have tuned these parameters through either analysis or simulation. The parameter $K_1$ is a fitting parameter with a distinct value for each access technology. The parameters $w_1$ and $w_2$ reflect the relative importance of wireless and backhaul loads in the calculation of the NAP quality. As an example, if congestion has a higher probability to occur in the wireless link than in the backhaul access then $w_1$ should be higher than $w_2$ (Table 3).

The *alpha* parameter can be used to adjust the choice of the more suitable NAP according to the terminal mobility pattern. As an example, if terminals are static (or nomadic) then this parameter can assume a value of 0.2. This means that the term of (5) associated with power is almost irrelevant for the final ranking result. Alternatively, the *alpha* parameter could assume higher values for mobile scenarios. The $POW_{THR}$ and

$QUAL_{THR}$ are sensitivity parameters for the variance of respectively received signal and the NAP quality metric. The *delta* is a hysteresis handover parameter; it is very important to guarantee globally the network stability and a low number of handovers, especially for high loaded heterogeneous networks (see Section 7.3), because it avoids the typical *ping-pong* that occurs during NAP selection. The QT (Quality Threshold) logically coordinates the distinct technologies in terms of balancing the load fairly among these, according to their available resources. QT also strongly affects flow admission behavior.

## 5. ANALYTIC STUDY

We aim to answer the following question: what long-term impact the current management proposal with a brokerage service would have on the operation of a heterogeneous wireless network access topology? To answer this, an analytical study in the context of a wireless network access topology covering a very busy public location was carried on. This location is covered by two distinct technologies, Wi-Fi and WiMAX, each one deployed by a distinct network provider.

This study uses two business models. The first model is between each network provider and its subscribers. It was assumed that each subscriber pays a fixed tariff to its network provider for each hour of connection, irrespective of the access technology used by that connection. The second one is between the network providers. When a provider has the problem of lacking network capacity to satisfy demand, it pays a fixed tariff for each hour of connection related to a subscriber successfully moved to the other network. Otherwise, the client is blocked and this situation is processed in the current study as a financial cost (i.e. churn) to the provider that has a contract with that client because unsatisfied clients normally change to an alternative provider offering a better connection service. It is assumed that each network provider has a market share of 50%. Our study incorporated real passenger loading statistics during a week at Euston railway station (Figure 4). This was obtained from the dataset published by Office of Railway and Road (ORR, http://orr.gov.uk/statistics/published-stats). We have also used some selected results (Table 4) discussed in Section 7 about typical values of how many terminals are attached to each AP / BS.

Tables 5-6 present further information about how our study was structured. In Table 5 there are two distinct deployment strategies for network provider A. The brokerage service enhances the management of the available network resources to offer wireless connectivity to the backbone network, meaning that a terminal flow can be moved between the access technologies or a flow can be blocked when no NAP is available with a quality higher than the specified QT value of 0.525. The parameter $p_a$ is the amount of money paid by each user to his subscription provider (in this case provider A), if the user terminal is connected, via either network A or network B, during a fixed time interval. It is a similar case for $p_b$. The parameter $p_h$ is the amount of money paid between providers, during a fixed time interval, in the case of a flow being exchanged between technologies. This payment acts as an incentive to a provider for accepting customers from others. The network of provider A is composed of 3 or 5 Wi-Fi Access Points (APs), depending on the deployment strategy of this provider (Table 5). Alternatively, provider B has an invariant network topology formed by a single WiMAX Base Station (BS) with a broad coverage. Table 6 shows that the cost of WiMAX equipment is higher than Wi-Fi. Another relevant cost is considered in the situation when a specific technology blocks flows. This potentially creates unsatisfied clients because they are poorly served by that technology. This implies churn of clients and the reduction on the associated operator's profit.

For the current study, distinct usage scenarios were studied (i.e. scenarios 1 and 2). In scenario 1 the brokerage service is disabled. The results of this scenario are shown in Figures 5 and 6. Analyzing first Figure 6, its generic trend suggests a degradation on the quality of connection service, essentially during peak times. Nevertheless, this degradation is slightly compensated if provider A chooses strategy 2. In addition, Figure 5 shows that if the brokerage service is disabled then provider B becomes the dominant provider in the market. This outcome is very unfair because provider B did not make any investment in upgrading its network infrastructure and offers a poor quality of service to its clients. This poor quality connection service should be factored into each provider's profit thus decreasing the profit as a long-term cost, representing the churn of unsatisfied customers. The trend of Figure 5 does not include this last cost.

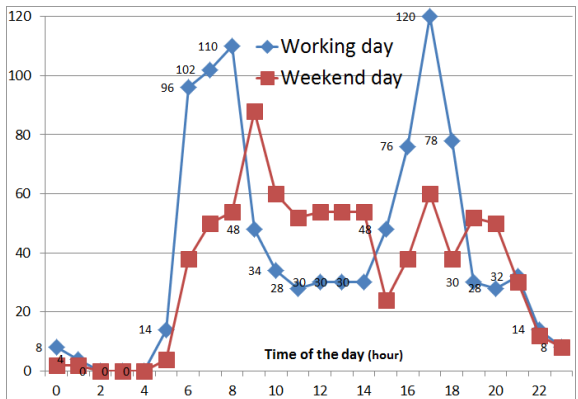

Figure 4 – The total number of customers requiring a connection at Euston station (UK) (scaled to the study network infrastructure) during typical working and weekend days

Table 4 - NAP Capacity

| NAP Type | Typical | Maximum |
|---|---|---|
| Wi-Fi AP | 8 clients | 11 clients |
| WiMAX BS | 24 clients | 33 clients |

Table 5 - Network Provider Strategies

| Strategy | $p_a$ | AP | $p_b$ | BS | $p_h$ |
|---|---|---|---|---|---|
| 1 | 0.45 u | 3 | 0.90 u | 1 | 0.68 u |
| 2 | 0.70 u | 5 | 0.90 u | 1 | 0.68 u |

Table 6 - Costs per access technology

| Cost | Wi-Fi | WiMAX |
|---|---|---|
| 1 Mbps of network infrastructure during a year | 600 u | 1200 u |
| Blocking a flow implies client´s churn | 1.35 u | 1.35 u |

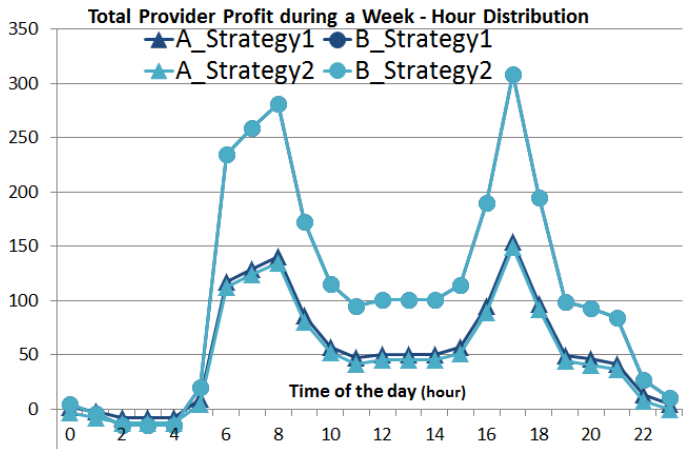

Figure 5 - Providers weekly profit per strategy (money units) with the broker disabled vs. hour of the day (scenario 1)

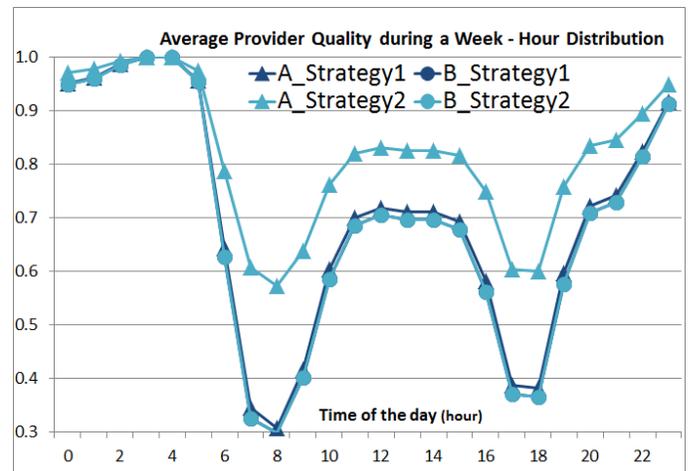

Figure 6 - Providers weekly quality per strategy with the broker disabled vs. hour of the day (scenario 1)

In the second scenario, the brokerage service is enabled. These results are shown in Figures 7 and 8.

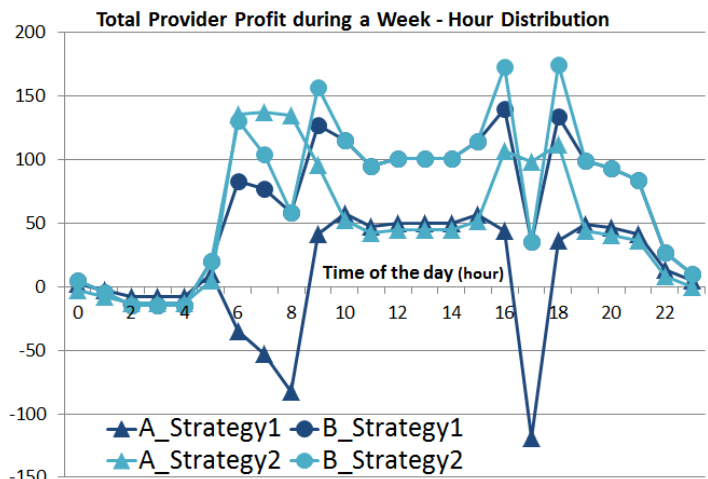

Figure 7 - Providers weekly profit per strategy (money units) with broker active vs. hour of the day (scenario 2)

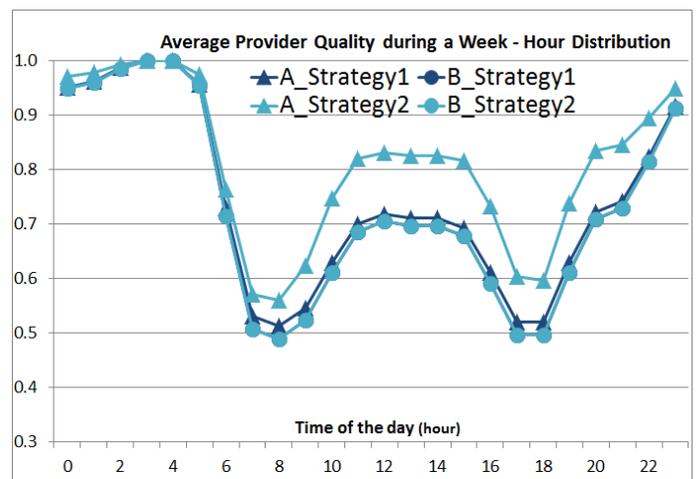

Figure 8 - Providers weekly quality per strategy with broker active vs. hour of the day (scenario 2)

Discussing the results visualized in Figure 7, the provider A becomes the provider with the dominant position in the market as that provider chooses the strategy 2. When the financial results of both scenarios are compared (i.e. Figures 5 and 7), as the brokerage service is enabled, it introduces fairness in the market, protecting the investment made by provider A and penalizing provider B, whom did not choose to upgrade its network capacity. Further, comparing the results of Figure 8 with the ones of Figure 6, we can conclude that the management of the entire network based on a brokerage service increases the service quality offered by all the access technologies, enabling operators to keep high reputation values in spite of very high loads.

The analytic study suggests that the brokerage service can help a provider to plan more accurately the right provision for its network, including the backhaul link. It also suggests that the brokerage service enables a globally improved usage of the available network resources.

## 6. IMPLEMENTATION

A significant number of changes have been made on a simulator [38] to evaluate comprehensively the major novelties of our solution. New functionality included is as follows: several broker agents at the network level deployed in both network and terminal sides (Figure 9); addition of quality metrics per technology to evaluate accurately the wireless and backhaul status; usage of a novel distributed algorithm combining both flow admission and network selection features that apply management policies to control the load; addition of the retransmission of MIPv6 redirect messages to implement a reliable make-before-break handover among distinct wireless technologies; incorporation of a proactive mechanism to evaluate the backhaul link of each technology; modification of used wireless protocols (i.e. Wi-Fi, WiMAX) to send the quality metrics to terminals; and the creation of a new event in the MIHF [15] API to refresh the terminal's agent running at the network layer with the quality metrics received by each terminal's MAC.

The brokerage service functionality is scattered among diverse network units (Figure 9). Each one of these units has a main module that is periodically processed with the following goals: specifies master or slave functionality for each unit; specifies the rate to measure the backhaul Round Trip Time (RTT) using ICMP (e.g. every 0.5 s); retransmits an ICMP request (e.g. up to a maximum of three retries) if the last transmitted request has not been confirmed by the selected Internet server within a specified time (e.g. 0.1 s); forces the slave unit to calculate quality metrics after a successful ICMP request message (re)transmission.

There are as many slave units as available access technologies. The slave units have the following goals: evaluate the status of the backhaul link using the previous explained measuring active method based on ICMP messages; associate the backhaul link quality with the wireless channel quality of each associated NAP to assess that NAP quality; calculate the technology quality as the average of all NAP quality values; evaluate the provider reputation as the average of the technology quality; reflect the technology status immediately on all the quality metrics, which are then disseminated to the related recipient network nodes (e.g. NAPs of the same technology and the master unit). Inside a specific network region or domain, there is a single master unit with the following tasks: receives the status of each technology from the associated slave unit; deploys a policy to prioritize the technologies; and updates each slave unit with the technology priority.

Some novel NAP functionality is examined as follows. Initially, the NAP calculates the local wireless quality according its load (i.e. number of connected terminals), and stores this value inside the MAC MIB. In addition, the NAP sends upstream the wireless quality metric stored at the MAC MIB towards the associated slave unit. The NAP also propagates downstream its quality metrics to terminals, using piggybacking in (periodic) wireless broadcast frames. Two distinct NAP implementations were built for IEEE 802.11 and IEEE 802.16. Each 802.11 Access Point (AP) uses the beacon frame to broadcast its quality to end-users devices. In addition, each AP can send a Probe Response frame, during a process of channel scanning, to announce the metrics directly to the terminal that initiated the scanning. This enables terminals to discover new APs using distinct channels. For 802.16, each Base Station (BS) uses the Mobile_Neighbour_Advertisement (MOB_NBR_ADV) frame to spread its own quality and the quality of neighboring BSs among terminals [39]. The BSs also exchange status via wired network, ensuring that WiMAX wireless channels are less overloaded with signaling than Wi-Fi ones.

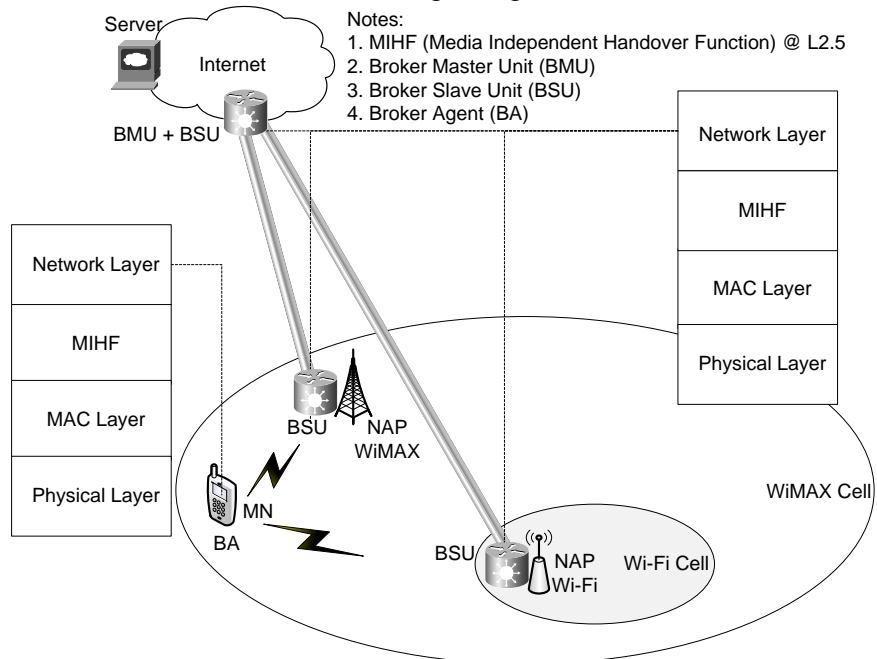

Figure 9 – Brokerage service implementation based on IEEE 802.21

This work assumes that each client uses a multimode terminal capable of simultaneously receiving link layer management frames from distinct technologies, each one announcing its own metrics. The terminal MAC layer associated with each technology extracts the metrics from the received messages at random time instances, joins to these metrics the related RSS and notifies the terminal's agent at the network layer through an intermediate layer, designated by MIHF [15], using a Link Update Event service, specially created for this purpose. These events keep the terminal's agent updated with the infrastructure status. After each update, the agent recalculates the NAP ranking list to select the more convenient NAP. This NAP should have a quality higher than the QT threshold. Otherwise the terminal is blocked.

Our implementation assumes the terminal should have enabled all the wireless interfaces to its agent can receive the technologies metrics. This could allow a terminal to send multiple flows in parallel through distinct technologies (i.e. multipath), enhancing that terminal's throughput. Nevertheless, this implementation has a limitation because the autonomy of the terminal's battery is reduced. So, we have declined the multipath functionality in our deployment.

## 7. EVALUATION

The current proposal was evaluated using a network domain similar to the shown in Figure 9. It is a correct topology to verify in a conceptual way if our brokerage service can successfully manage at the network edge several heterogeneous wireless technologies, e.g. Wi-Fi and WiMAX. WiMAX covers a larger area than Wi-Fi, as Table 7 shows. On one hand, WiMAX is the only technology available to the terminals located at the edge of the testing area. On the other hand, the terminals located inside the central area can detect both technologies but with diverse signal strengths depending on their location. During each experiment, the number of terminals located within the coverage area that requires for network connectivity is changing. The following aspects have been evaluated: i) enhancement on the wireless quality of each accepted flow via a distributed management algorithm (Section 7.1); ii) how the number of accepted flows in the wireless link is adjusted to constraints on the wired backhaul? (7.2); iii) how the distributed algorithm behaves after consecutive flash crowds? (7.3); iv) centralized policy vs. distributed flow control (7.4).

The evaluation used a modified implementation of a simulator [38]. Further information about the network topology is available in Table 7. Table 8 shows the default configuration parameters (i.e. policy) used by the brokerage service. The Quality Threshold (QT) controls at the terminals the user admission and NAP selection. The weights ($w_1$, $w_2$) assume that congestion occurs with a higher probability at the wireless media.

Each scenario was simulated via ten iterations. The simulation time of a single iteration was roughly five minutes. During this time, the network load was increasing with a new flow per second, with the first flow being added at nine seconds after system's initialization. Each flow had a Constant Bit Rate (CBR) of 320 Kbps. All flows were originated at the server connected to the wired network and destined to multimode terminals, a flow per terminal. Each flow has an airtime path with a single hop. The main goal of our evaluation was to guarantee a maximum quality to each connected flow. In this way, each connected flow should be served with a rate tending to 320 Kbps. All obtained results have a confidence of 95%. The CBR traffic was used because it was suitable to evaluate the major functionalities of our solution. As future work, it would be interesting to study the brokerage system with other traffic types; it could be necessary to select some different parameters (i.e. policies of Table 8) to control the flows in an efficient way.

Table 7 – Simulation Parameters

| Node | Coverage (m) | Speed (m/s) | Mobile | X_src | Y_src | X_dst | Y_dst |
|---|---|---|---|---|---|---|---|
| BS | 1000 | - | No | 1000 | 1300 | 1000 | 1300 |
| AP1 | 20 | - | No | 995 | 1000 | 995 | 1000 |
| AP2 | 20 | - | No | 1005 | 1000 | 1005 | 1000 |
| Static Terminal | - | 0 | No | 1000 | 999 | 1000 | 999 |
| Mobile Terminal | - | 1(1) | Yes | 900 (even), 1100 (odd) | 999 | 1000 | 999 |

Note (1): Each pair of consecutive mobile terminals starts its movement with a time interval of 5 s, starting with the first pair at 10s.

Table 8 – Algorithm Parameters

| Terminal Agent | | Broker Slave Unit | |
|---|---|---|---|
| Power Threshold (Pow_thr) | Quality Threshold (Qual_Thr, QT) | Weight for Backhaul Quality ($w_2$) | Weight for Wireless Quality ($w_1$) |
| 7e-9 | 0.525 | 0.2 | 0.8 |

## 7.1. Wireless Scenarios

The Table 9 lists five distinct testing scenarios (A to E). During all these tests, the wireless media is gradually congested with data traffic up to a maximum load of 80 terminals. The scenario A is the only case where the brokerage service is disabled. All the terminals (i.e. clients or flows) in scenarios A, B and C are static, and the remaining scenarios have also a variable number of mobile terminals. In scenario A, all the terminals initially connect to the network backbone via a default access technology (Figure 10.a) and stay connected in spite of the degradation of their connection quality (Figure 10.b), reaching a maximum degradation of 37% at 90s. This best-effort connection service occurs because the broker functionality is disabled, and consequently the network infrastructure cannot control the connection quality offered to each flow. In this way, almost all the terminals choose WiMAX (Figure 10.a, 15 nodes at t=30s and 80 nodes at other instances of time) because WiMAX typically has a stronger and broader coverage than Wi-Fi.

Table 9 – Wireless Scenarios (A, B, C, D, E)

| Scenario | Brokerage Service Enabled | Qual_Thr | Static Terminals | Mobile Terminals |
|---|---|---|---|---|
| 0% Mobile (A) | No | 0.525 | 80 | 0 |
| 0% Mobile (B) | Yes | 0.525 | 80 | 0 |
| 0% Mobile (C) | Yes | 0.725 | 80 | 0 |
| 20% Mobile (D) | Yes | 0.525 | 64 | 16 |
| 60% Mobile (E) | Yes | 0.525 | 32 | 48 |

Scenario B is the first case where the brokerage service is enabled. This service measures the instantaneous quality of the two technologies enhancing the user admission and network selection made in each terminal, which is supported by quality metrics and management policies. This allows load balancing and offloading between technologies (Figure 10.a) controlled by user admission across the technologies, depending on the maximum acceptable load of each technology. In this case, a single WiMAX BS can support more users than the aggregation of two Wi-Fi APs. The benefits of using a brokerage service are easily visible when in Figure 10.b both results of scenarios A and B are compared. In fact, it is visible a quality increase of 37.5% (from 62.5% to 100%, where 100% represents a flow rate of 320Kbps) at the end of each test. In parallel, the total number of flows also decreases 59% in Figure 10.a (t=150s, from 80+0 down to 27+20) because user admission blocks flows to satisfy the policy that imposes a minimum quality threshold (QT) of 0.525. As QT is increased to 0.725 (scenario C), all attached flows tend more quickly to a perceived quality of 100% (Figure 10.b) because less flows stay connected (Figure 10.a) when compared with scenario B (t=270s, 46 vs. 23 flows). This is justified by a higher value of QT (0.525 vs. 0.725). See also our results in Section 7.4 that further discusses this topic.

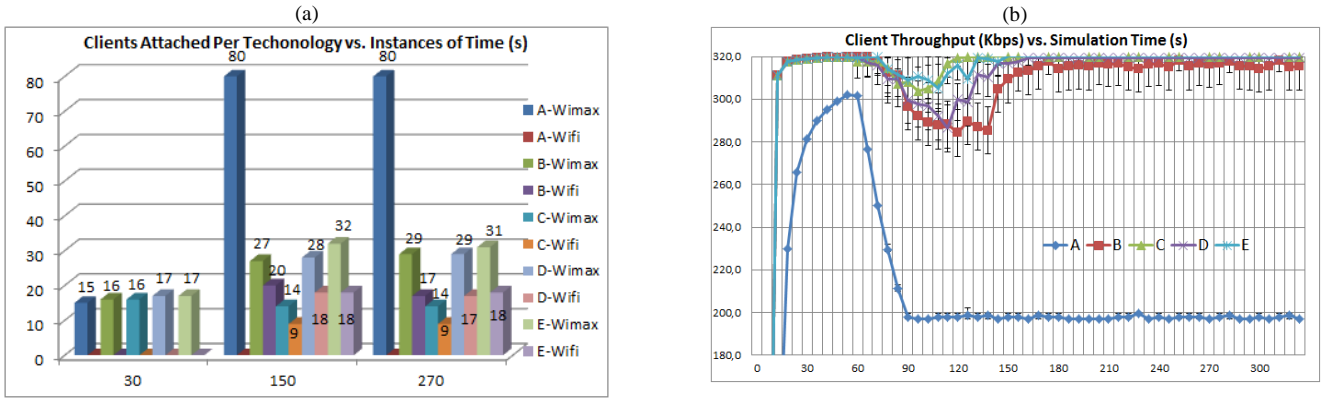

Figure 10 – Evaluation results for scenarios A, B, C, D and E vs. simulation time (s) (a) instances of time; b) continuous time).

Figure 10.b also presents the results of mobility scenarios (D and E). In these scenarios, a flow handover can result from the combined effect of two distinct and simultaneous dynamics: load variation and terminal movement (all the mobile terminals have the same speed of 1 m/s). The mobility scenarios of Figure 10.b show a better performance in terms of throughput than the static-only scenarios despite having to cope with the movement of the terminals. The mobile scenarios seem apparently less complex to manage by the distributed algorithm running in each terminal because at the beginning of each mobility scenario the terminals located in the central area covered by the two technologies are fewer than the ones of static-only scenarios. This occurs because the mobile terminals are initially located far away from the central area, where only WiMAX coverage is available. In this way, as the distributed algorithm is initially running between only a few concurrent agents because the mobile terminals have not yet arrived at the central area, the algorithm in execution at each terminal converges faster. This means the algorithm is able to initially balance the load of the static terminals between the technologies in a way that the number of blocked flows is lower than in the static scenario, to satisfy the same conditions of the policy that is being targeted.

To enhance our evaluation in mobile scenarios, it was also used an open-source Java tool [40]. This creates realistic mobility scenarios that have been exported to our simulator. The selected mobility scenario was *RandomWaypoint* in which the nodes are distributed randomly over the simulation area. After waiting for a constant pause time, each node chooses a (random) waypoint and moves towards it with a speed randomly chosen from a specified interval. After arriving to this waypoint, the node pauses for a constant time and chooses the next (random) waypoint. To generate a *RandomWaypoint* scenario with 80 nodes randomly moving for 300 seconds within a topological area of 25x25 meters, the following command was used: "*bm -f mob_scn RandomWaypoint -d 300 -n 80 -x 26 –y 26 -o 3 -h 1.2 -l 0.8 -p 60*" [40]. Table 10 lists the simulation parameters of this scenario. In spite of mobility in all the terminals, the distributed management algorithm of the brokerage service offered to each connected flow an average throughput over 94% of its maximum value of 320 Kbps, with a limited number of handovers (46) between technologies.

The current results illustrate the positive impact on the flow quality when the brokerage service is enabled within a heterogeneous network infrastructure. Further, this positive impact can be easily tuned, using a centralized and configurable approach to choose a more suitable set of management policies based on parameters and metrics, depending on the usage scenario.

Table 10 - Scenario Simulation Parameters

| Node | Coverage (m) | Speed (m/s) | x | y |
|---|---|---|---|---|
| BS | 1000 | - | 326 | 10 |
| AP1 | 20 | - | 8 | 26 |
| AP2 | 20 | - | 18 | 26 |
| Terminal | - | 0.8 - 1.2 | (1) | (1) |

Note (1): Follows the generated mobility RandomWaypoint scenario

## 7.2. Backhaul Scenarios

We have also studied how distinct provision levels of backhaul bandwidth could be managed by the brokerage service (Table 11). The initial wireless traffic load was always 12.8Mbps. The broker slave unit has been adjusted to a more convenient policy, with 0.5 for both weights of backhaul and wireless quality. In this way, a load increase in either the backhaul or the wireless channel reduces the NAP quality at the same proportion. Scenario F corresponds to a well-provided backhaul for the initial wireless load. In opposition, Scenarios G and H are both related to a backhaul lacking capacity for that load.

Table 11 – Backhaul Scenarios (F, G, H)

| Scenario | WiMAX Backhaul Capacity (Mbps) | Wireless Load (Mbps) | Qual_Thr | ($w_1$, $w_2$) |
|---|---|---|---|---|
| Overprovisioned (F) | 15 | 12.8 | 0.525 | (0.5, 0.5) |
| Underprovisioned (G) | 5 | 12.8 | 0.525 | (0.5, 0.5) |
| Underprovisioned (H) | 5 | 12.8 | 0.725 | (0.5, 0.5) |

Scenario F results show the handovers are policy-managed in a way that both technologies share the wireless load because the quality visualized in Figure 11.a, after t=90s, stays around 56.5%. This is a very important result because it illustrates handover-based load distribution obtained by running the same management algorithm across distinct access technologies and controlled by centralized policies. In addition, in Figure 11.b no congestion is visible in the WiMAX backhaul because it is overprovisioned. As the backhaul quality (Figure 11.b) is combined with the wireless quality (Figure 11.a), using the weights visualized in Table 8, both technologies stay around the same quality, i.e. 78.5% (i.e. 0.57*0.5 + 1*0.5=0.785), a value slightly above the QT of 52.5%. This difference of 26% between the quality of each technology and the QT can be justified by several factors namely the propagation delay of the quality metrics dissemination through the system and the hysteresis *delta* parameter (Section 4.6).

Results of scenario G suggest that as the WiMAX backhaul has not enough capacity to support 40 flows (i.e. 12.8Mbps) with a satisfactory quality then the distributed algorithm actively drops back the number of flows attached to WiMAX. In this way, some flows are offloaded to Wi-Fi and others are blocked. This is why, in scenario G, there are more flows attached to Wi-Fi (Figure 11.c, 21 vs. 17) and less flows connected to WiMAX (Figure 11.c, 17 vs. 23). The results from Figure 11.d (scenario G) suggest that a value of 0.525 for the QT is too low for ensuring a maximum quality for the flows connected via WiMAX when the backhaul is slightly congested. This requires more flows to be blocked, increasing the quality of the flows that stay connected. This explains why the results of scenario H with a higher QT show in Figure 11.d a maximum flow throughput; in Figure 11.c a decrease on the attached number of flows (from 38 to 26); and in Figure 11.b an increase on the WiMAX backhaul quality (from 0.86 to 1). The results suggest that the number of accepted flows in the wireless medium was adjusted to the backhaul constraint. Further, traffic offloading is always made from the access technology with the backhaul congested to other technologies with enough network resources, including backhaul capacity. Finally, these significant achievements are obtained through a distributed algorithm executed in each terminal agent and controlled by the QT policy.

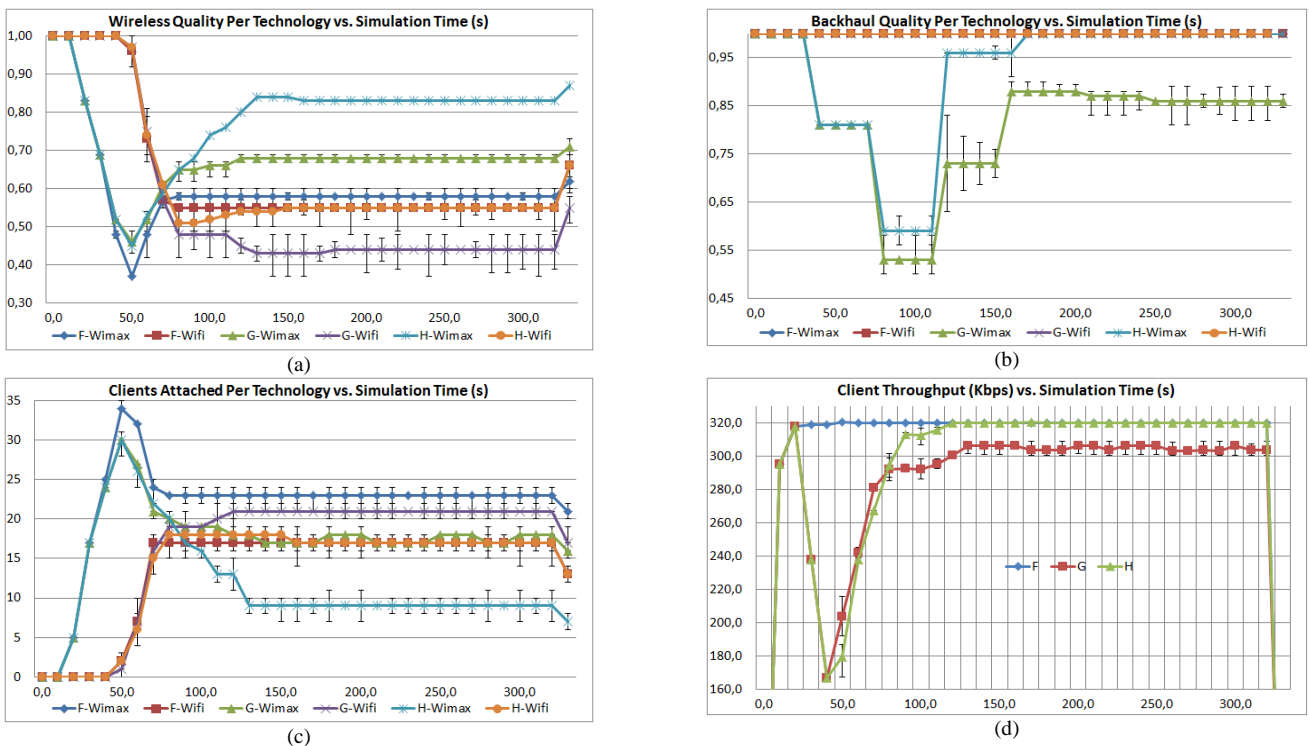

Figure 11 – Evaluation results for scenarios F, G and H vs. simulation time (s)

*7.3. Algorithm Stability*

Our distributed management algorithm attempts to find a final stable distribution of flows among heterogeneous access technologies. In this way, we have also studied the stability of that final flow distribution. The network infrastructure has initially a total traffic load of 25.6Mbps (80 flows). This initial load is enough, in the test scenario, to overload the wireless media, forcing the distributed algorithm to counteract the negative load impact on its own quality. In addition, the QT value has a value of 0.525 and all the terminals are static during the entire duration of this evaluation. Further, the simulation time was increased to 900s. In the current evaluation scenario (i.e. scenario I), after the distributed management algorithm found a viable distribution of the initial load between the available network resources of the two access technologies, some flash crowds (40 new flows in each flash crowd) were generated, at three distinct occurrences, for testing algorithm stability. The obtained results are shown in Figures 12-15.

Figure 12 shows that WiMAX has accepted, at the initial phase of each flash crowd situation, i.e. 260s, 460s and 660s, 21, 7 and 2 new flows (clients), respectively. Nevertheless, after several seconds at each congestion occurrence, most of the flows initially accepted by the WiMAX technology are disconnected to compensate for the drop in the quality of that technology. In spite of the extra traffic load, there is not, at the occurrences of extra load, any noticeable degradation on the averaged individual client (flow) throughput (Figure 13), with a very low number of handovers (below 17). The negative impact on both packet loss (Figure 14) and inter-arrival packet delay (Figure 15) are limited, and sometimes unnoticeable. All these performance results suggest low signaling network overhead to mitigate the negative impact of sequential flash crowds.

In short, after the initial distribution of load between the technologies has been found, the management algorithm shows flexibility in terms of accepting additional load, without affecting the service quality offered to each connected terminal, in spite of severe and consecutive network overloads.

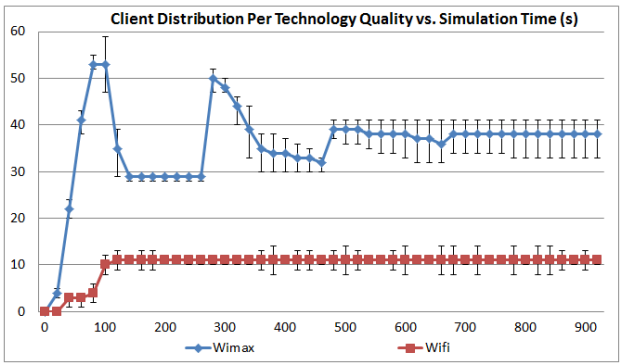
Figure 12 – Flows per technology vs. simulation time (s) for scenario I

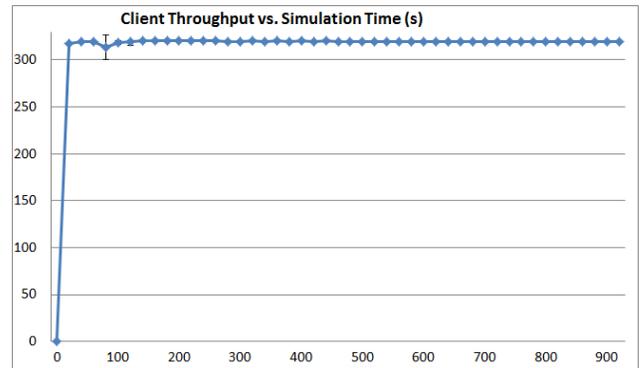
Figure 13 – Flow throughput (Kbps) vs. simulation time (s) for scenario I

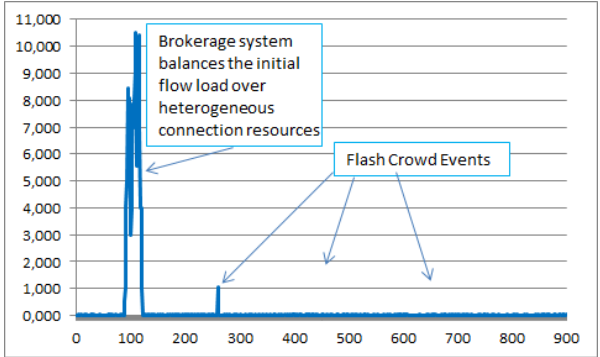
Figure 14 - Lost packets per flow vs. simulation time(s) for scenario I

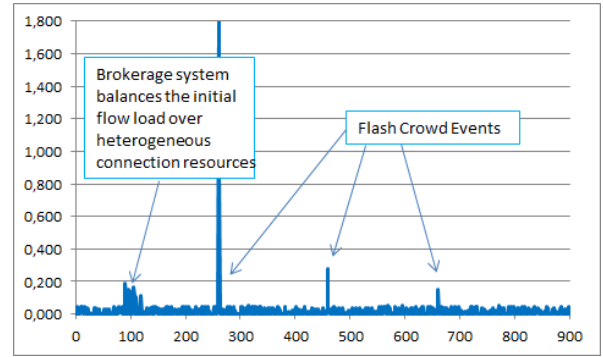
Figure 15 - Inter-arrival packet delay(ms) vs. simulation time(s) for scenario I

### 7.4. Operation like SDN

We have also evaluated the scenario J ($w_1$=0.8, $w_2$=0.2) with several values for wireless loads and QT. The first comment on these results (Figure 16) is that the distributed algorithm requires more time to converge, for high wireless loads, to the final system status. Our second observation is that using a more rigid (protective) policy on the flow quality (QT=0.725) it implies that fewer flows stay connected at the end of each testing scenario, when compared with analogous results for a more relaxed (flexible) quality policy, i.e. QT=0.525. All these conclusions suggest that the brokerage solution operates similarly to what is expected from SDN (centralized logical control vs. distributed data routing/switching). QT can also be seen as a very effective Northbound SDN policy to rule flows admission and their subsequent management.

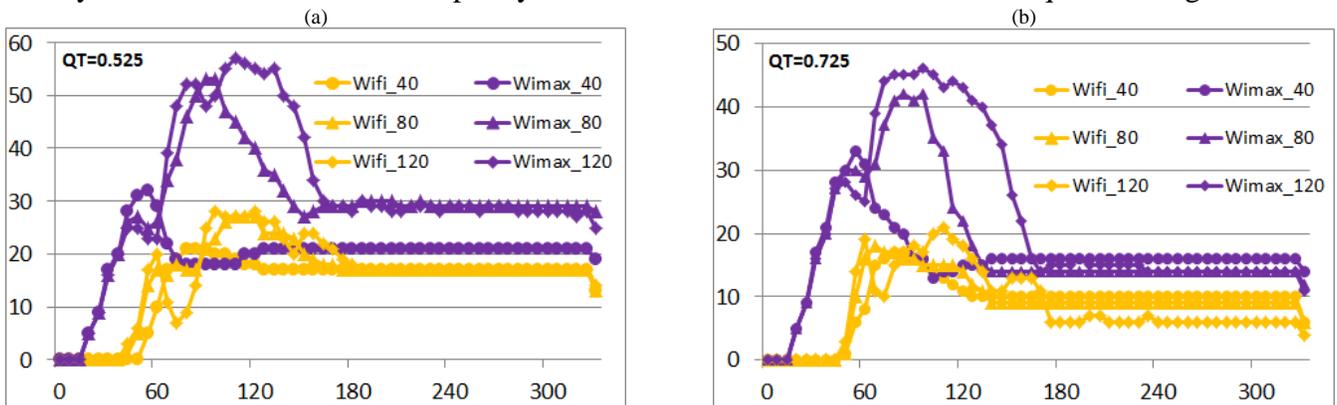
Figure 16 – Flows attached per access technology vs. simulation time (s) for scenario J

Finally, Table 12 summarizes the most relevant results of our evaluation. We argue that the obtained results are completely independent of the used access technologies. During our current work, we have selected WiMAX and Wi-Fi but we are strongly convinced that similar results can be obtained with distinct access technologies (e.g. cellular vs. Wi-Fi). For this, some minor adaptations would be needed: the methodology to measure the wireless quality and the backhaul quality, or how the metrics are processed.

Table 12 - Evaluation Main Results

| Evaluation Scenario | Main Result |
|---|---|
| Wireless Overload | Quality offered to each connected flow increases 37.5% when the brokerage service is enabled; this service is still very effective with terminal mobility, including the well-known Random Waypoint pattern |
| Constrained Backhaul | Number of flows accepted in wireless medium is perfectly adjusted to the backhaul provisioning and load balancing/offloading is made amongst available access technologies |
| Algorithm Stability | Distributed algorithm executed at each terminal agent converges to a stable number of connected flows through each technology in spite of consecutive flash crowds, with a low network overloading |
| Operation like SDN | Through a simple change on a single and centralized system configuration parameter (i.e. QT) is possible to control in a distributed way the logics of flow admission and connection over a completely heterogeneous networking access infrastructure |

## 8. CONCLUSION

Our novel hybrid brokerage service deals with congestion in a mobile scenario, including the backhaul access. This solution enhances flow admission and network selection among distinct access technologies, supporting a specific level of connection quality in spite of flash crowds. Using real data about the network usage, an analytic study highlighted the positive outcome of this novel solution on the long-term financial profit of each network provider, when operators cooperate enabling the roaming of clients. This also means that each flow gets the best possible connection quality in spite of congestion, and the market fairness is ensured. The brokerage service is deployed at the network level and is completely independent of the link layer technologies. As a proof of concept, our evaluation clearly shows that this service enhances the operation of a heterogeneous wireless infrastructure in realistic scenarios with significant loading and node mobility, using a distributed, multicriteria, and policy-controlled algorithm. It also produces stable management decisions in spite of consecutive flash crowds, with a low network overloading. The usage of the brokerage service in scenarios with femtocells seems a very interesting future research direction. In addition, if necessary, the main innovations of our proposal can be migrated to an environment combining both SDN/NFV [20-23, 41] and Wireless Networking for Moving Objects (WiNeMO) [5].

ACKNOWLEDGMENTS

This work was supported by *Fundação para a Ciência e a Tecnologia* (FCT) under Grant SFRH/BD/28193/2006.